\documentclass[epj]{svjour}
\usepackage{graphics}
\usepackage{amsfonts}
\usepackage{amsbsy}
\usepackage{mathrsfs}
\usepackage{amsmath}
\usepackage{amssymb}
\usepackage{slashbox}
\usepackage{graphicx,epsfig}
\begin{document}
\title{Flavor changing neutral couplings for leptons}
\author{E. A. Paschos\thanks{\email{paschos.e@gmail.com}}
}                     
\institute{Department of Physics, TU Dortmund, D-44221 Dortmund, Germany}
%
%
\abstract{The article reviews the conditions for lepton-flavor conservation in leptonic interactions. The conditions are similar to those developed in the quark sector. For neutrinos the mass eigenstates are unitary superpositions of flavor states which together with small mass differences produce neutrino oscillations. A new property for neutrinos is the fact that their masses are very small suppressing their couplings to neutral scalar fields.  The results are demonstrated in several models.
%
} 
\authorrunning{E. A. Paschos}
\titlerunning{Flavor changing neutral couplings for leptons} \maketitle
%
%
\section{Introduction}
In gauge theories the structure of the currents is determined by the algebra of the group~\cite{Glashow:1970gm,Paschos:1976ay,Glashow:1976nt,Paige:1977nz}. One manifestation of the algebra is the introduction of restrictions on the group assignments for quarks and leptons due to the absence of flavor changing neutral currents (FCNC). Flavor changing neutral couplings have not been observed in lepton decays~~\cite{FCNC exp} and extensions of the standard model predict~\cite{Cheng:1980tp} a decay rate for $\mu\rightarrow e+\gamma$ which is very small. The extensions require additional neutral currents or right-handed neutrinos which generate neutrino masses and mixings. The suppression of lepton changing decays is related to the representation assignments of leptons and to the small masses for neutrinos. 

For leptons, however, flavor changing couplings must be present in neutrino mass matrices in order to produce neutrino oscillations. Thus it is interesting to review and compare flavor changing neutral interactions for quarks and leptons. In the standard model and its minimal extensions special properties arise. This motivates a study of properties required by FCNC for leptons in the standard model and also in models with extended groups, like $SU(2)_L\times U(1)_Y\times U(1)_H$. 

In gauge theories based on $SU(2)_L\times U(1)$ the following assumptions are introduced~\cite{Paschos:1976ay,Glashow:1976nt,Paige:1977nz}:

(1) Direct neutral current couplings conserve lepton flavors to order $G_F$.

(2) Neutral couplings induced by one-loop radiative corrections conserve lepton flavors to order $\alpha G_F$.

Applying them to leptons leads to analogous rules that have been developed for quarks (see remarks in summary).

There is a third requirement:

(3) The couplings of each neutral Higgs meson conserves all quark flavors to lowest order~\cite{Glashow:1976nt}. The situation for the third requirement is different for leptons. The couplings of neutral Higgs mesons should conserve lepton flavors between charged leptons, but produce lepton flavor transitions for neutrinos. In the traditional approach of generating lepton masses, through Yukawa interactions, the very small masses of neutrinos automatically suppress flavor-changing couplings of neutrinos to scalar particles. The physical neutrino states are unitary superpositions of lepton flavors which together with the small mass differences produce the oscillations.

The plan of the article is as follows. In section~2, I review how the requirements for FCNC of leptons are implemented in the standard model. It is stated there that the very small masses of neutrinos suppress flavor-changing interactions of neutrinos with scalar particles. In section~3, I extend the analysis to theories based on a larger group $SU(2)_L\times U(1)_Y\times U(1)_H$. The first is a model with a $\tau_{3R}$ or $(B-L)$ symmetry and the second includes 2HDM.

\section{Sterile neutrinos}

The oscillation phenomena require neutrinos to have masses and, in addition, their interactions must change lepton flavors. Thus introducing three right-handed neutrinos as singlet particles of $SU(2)_L$ allows the following Yukawa interactions

\begin{equation}
\mathcal{L}_Y=Y^{ij}_e\bar l_iHe_{Rj}+Y^{ij}_\nu\bar l_i\tilde H\nu_{Rj}+h.c. \label{eq:yui1}
\end{equation} 
with $l_i$ the left-handed doublets with lepton flavor $i$, $e_{Ri}$ and $\nu_{Ri}$ are right-handed singlets and $H(x)$ the standard Higgs doublet. In the flavor basis the states are denoted with a prime $l_{Li}=\left(\begin{array}{c}
\nu'_i \\ e'_i\\
\end{array}
\right)_L$ and $e'_{Ri}$ and $\nu'_{Ri}$. When the Higgs acquires a VEV, denoted by $v$, masses are generated for neutrinos and charged leptons

\begin{equation}
M^{\nu}_{ij}=\bar \nu'_{Li}M_{ij}\nu'_{Rj}\,\,\,\,\,\,\,\text{with}\,\,\,\,\,\,\,\,M_{ij}=Y_\nu^{ij}v. \label{eq:yui2}
\end{equation} 

They are diagonalized by unitary matrices acting to the left $U_L=U$ and to the right $U_R$. Similarly the mass matrices for charged leptons are diagonalized by corresponding matrices $V_L=V$ and $V_R$. Initially, the charged current is written in terms of flavor states and after their replacement with mass eigenstates the charge current takes the form

\begin{equation}
J_\mu=\bar \nu_{Lj}\gamma_\mu U^\dagger_{\alpha j}V_{\alpha i}e_{Li}\label{eq:yui3}
\end{equation}
with a summation over repeated indices. We can write the doublet as 

\begin{equation}
\Psi_L=\left(\begin{array}{c}
W_{ij}\nu_{Lj} \\ e_{Li}\\
\end{array}
\right)\,\,\,\,\,\,\,\text{with}\,\,\,\,\,\,\,\,W_{ij}=U^\dagger_{\alpha i}V^e_{\alpha j}, \label{eq:yui4}
\end{equation} 
where $e_{Li}$ and $\nu_{Li}$ are physical states. The upper component is a coherent superposition of physical eigenstates, which in their time development mix among themselves. By construction the number of charged and neutral leptons are equal. In addition the commutator of the charged currents produces neutral currents which conserve lepton flavors.

Some one-loop radiative corrections to order~$\alpha G_F$ are proportional to~\cite{Paschos:1976ay,Glashow:1976nt,Paige:1977nz}

\begin{equation}
T_{\pm}T_{\mp}=T^2-T_3^2\pm T_3 \label{eq:yui5}
\end{equation}
which restrict $T_L^2$ and $T_{L3}$ of leptons.
Flavor changing corrections to leading order vanish, when we sum over all intermediate states. Smaller terms of order

\begin{equation}
\alpha G_F\left(\frac{m_i}{m_W}\right)^2 \label{eq:yui6}
\end{equation}
survive. These properties restrict the assignments of leptons into representations of the group and guarantee the universality of the couplings~\cite{Paschos:1976ay,Glashow:1976nt}. In the next section I discuss the situations that arise when the group is enlarged with a $U(1)_H$ factor.

There is a third requirement demanding that couplings of quarks to Higgs scalars conserve flavor number to lowest order. To fulfill this demand a rule is introduced that quarks of the upper charge couple to one Higgs scalar and quarks of lower charge to a different scalar. In this case Dirac mass terms and neutral Higgs couplings become diagonal at the same time.

The situation may be different for leptons. We need flavor changing couplings for neutrinos in order to produce neutrino oscillations. However, the very small masses of neutrinos restrict lepton flavor changing interactions of neutrinos to be extremely small. I demonstrate this property with an example.

\begin{figure}
\centering
\includegraphics[width=0.3\textwidth]{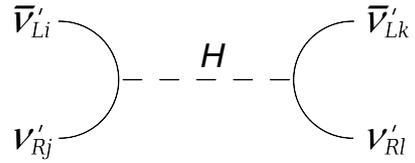}
\caption{Neutrino interactions with an exchange of Higgs.}\label{fig_1}
\label{fig:1}
\end{figure}

In the extension of the standard model with right-handed neutrinos there is the coupling $\nu'_{Li}+H\rightarrow \nu'_{Rj}$. The exchange of the Higgs produces interactions

\begin{equation}
\mathcal{L}_{int}=Y_{ij}\bar \Psi'_{\nu_{Li}}\tilde H\nu'_{Rj}\rightarrow Y_{ij}(H'_0+v)\bar \nu'_{Li}\nu'_{Rj} \label{eq:yui7}
\end{equation}
responsible for the process in figure~\ref{fig_1}. The couplings are matrices $Y^{\nu}_{ij}=m_\nu/v$ with $m_\nu$ a typical neutrino mass. The mass $m_H=125.9$~GeV and for neutrinos $m_\nu\sim10^{-9}$~GeV which determine the amplitude

\begin{equation}
S_{fi}=\frac{(Y_{ij})^2}{m_H^2}\approx\left(\frac{m_\nu}{v}\right)^2\frac{1}{m_H^2}\eqsim 10^{-27}~\text{GeV}^{-2}.\label{eq:yui8}
\end{equation}
This amplitude is very small compared to the Fermi interaction $S_{\text{Fermi}}=G_F\approx10^{-5}~\text{GeV}^{-2}$. Even if we replace $m_\nu$ by a mass of 100~eV~\cite{c_bounds} for neutrinos, we obtain

\begin{equation}
S_{fi}=\left(\frac{m_\nu}{v}\right)^2\frac{1}{m_H^2}= 10^{-23}~\text{GeV}^{-2}.\label{eq:yui9}
\end{equation}
In equation~\eqref{eq:yui4} the neutrinos are unitary superpositions of mass eigenstates. Neutrino masses are very small and the above argument guarantees that interactions with the standard Higgs are sufficient suppressed to be consistent with observations.  Bounds for lepton flavor transitions are obtained from the branching ratios~\cite{FCNC exp}

\begin{equation}
Br(\mu\rightarrow e\gamma)\leq5.7\times10^{-13}\label{eq:yui10}
\end{equation}
and

\begin{equation}
Br(\tau\rightarrow \mu\gamma)\leq4.4\times10^{-8}.\label{eq:yui11}
\end{equation}

\section{Extensions to a larger group}
\subsection{Model with one Higgs doublet\label{sub31}}
A popular extension of the standard model contains a multiplicative group factor $U(1)_H$~\cite{Fayet:1989mq,Appelquist:2002mw,Lindner:2018kjo}. These models attracted more attention~\cite{Boehm:2020ltd,Bally:2020yid,AristizabalSierra:2020edu,Lindner:2020kko,Khan:2020vaf,Alikhanov:2019drg,Alikhanov:2019drg1} after the publication of the XENON1T result. Many articles extend the standard model by introducing a new gauge boson associated with the $U(1)_H$ group and several scalar fields.

In the simplest realization, Dirac masses are generated by the standard model Higgs $H(x)$ and a singlet field $\phi(x)$ which generates Majorana mass term~\cite{Alikhanov:2019drg,Alikhanov:2019drg1}. Since only one scalar contributes to Dirac masses, the physical states are unitary superpositions of flavors and the neutral currents conserve lepton flavors, as described in the previous section.

The next possibility couples quark pairs to bilinears of scalar fields as follows~\cite{Dutta:2019fxn,Dutta:2020enk} with $H$ the SM Higgs and $\phi$ a new scalar being singlet of $SU(2)_L$

\begin{eqnarray}
\mathcal{L}_{Y}=-\frac{\lambda_u}{\Lambda}\tilde H\phi^*\bar Q_Lu_R-\frac{\lambda_d}{\Lambda}H\phi\bar Qd_R\nonumber\\-\frac{\lambda_\nu}{\Lambda}\tilde H\phi^*\bar l_L\nu_R-\frac{\lambda_e}{\Lambda}H\phi \bar l_Le_R. \label{eq:yui12}
\end{eqnarray}
The couplings $\lambda_u$, $\lambda_d$, $\lambda_\nu$ and $\lambda_e$ are square matrices contracted. A typical mass matrix is $m^{ij}_\nu=\dfrac{\lambda^{ij}_\nu}{\Lambda}vV$. The introduction of VEVs $v$ for $H$ and $V$ for $\phi(x)$ generates masses for fermions. The couplings are now the following

\begin{eqnarray}
\frac{\lambda_\nu}{\Lambda}VH\bar\nu_L\nu_R=\frac{m_\nu}{v}H\bar\nu_L\nu_R, \label{eq:yui13}
\end{eqnarray}

\begin{eqnarray}
\frac{\lambda_\nu}{\Lambda}v\phi\bar\nu_L\nu_R=\frac{m_\nu}{V}\phi\bar\nu_L\nu_R, \label{eq:yui14}
\end{eqnarray}
with typical neutrino masses given by $m_\nu=\dfrac{\lambda_\nu}{\Lambda}vV$.

The amplitudes generated through the exchange of the standard model Higgs have the same magnitudes as those computed in equations~\eqref{eq:yui8}--\eqref{eq:yui9}. In the second amplitude the scalar $\phi$ is exchanged producing

\begin{equation}
A_2=\left(\frac{m_\nu}{V}\right)^2\frac{1}{m_\phi^2}\simeq 10^{-12}~\text{GeV}^{-2}\label{eq:yui15}
\end{equation}
computed for $V=10$~MeV and $m_\phi=100$~MeV~\cite{Dutta:2019fxn,Dutta:2020enk}. Both amplitudes are smaller than the Fermi interactions, $G_F$. The model has the particular property that the couplings of neutral scalars to neutrinos are proportional to the mass matrix.

\subsection{Two Higgs doublet models \label{sub32}}
There are several extensions of the SM with two Higgs doublets (2HDM). We denote the two scalar fields by $H_1(x)$ and $H_2(x)$. They are introduced in order to avoid FCNC with the fermions classified in special ways under the new group $U(1)_H$. There are two types of models. In the simplest model of type-I, all standard model fermions receive masses only from the VEV of $H_1(x)$. Thus all standard model fermions preserve their previous properties.

An additional requirement is the cancellation of triangle anomalies. There are several $U(1)_H$ charge assignments that are anomaly free. Table~\ref{tab_y} shows charge assignments for three models which satisfy the conditions summarized in~\cite{Appelquist:2002mw}. The model denoted as $U(1)_R$ has a $\tau_{3R}$ symmetry and is very similar to a model with a simpler Higgs structure~\cite{Alikhanov:2019drg,Alikhanov:2019drg1}. An explicit cancellation of anomalies is described in~\cite{Alikhanov:2019drg,Alikhanov:2019drg1}.

\begin {table}
\begin{center}
\begin{tabular}{ |c|c|c|c| }
  \hline
\backslashbox{State}{Model}& $U(1)_R$ & $U(1)_{B-L}$ & $U(1)_Y$ \\\hline
   $\nu_L$, $e_L$ & 0 & -1 & -1/2 \\
	    $\nu_R$ & 1 & -1 & 0\\
            $e_R$ & -1 & -1 & -1 \\
     $u_L$, $d_L$ & 0 & 1/3 & 1/6 \\
     $u_R$ & 1 & 1/3 & 2/3 \\
     $d_R$ & -1 & 1/3 & -1/3 \\
		 \hline
		 $H_1$ & 1 & 0 & 1/2 \\ 
  \hline
\end{tabular}
\caption {Charge assignments for the 2HDM of type-I~\cite{Ko:2012hd}.}\label{tab_y}
\end{center}
\end {table}

In the type-II models the two scalar doublets carry different charges of $U(1)_H$~\cite{Ko:2012hd,Ko:2015fxa}. Only one scalar couples to leptons with the same electric charge. $H_1(x)$ couples to neutrinos and $H_2(x)$ to leptons with lower charge. 

\begin{equation}
\mathcal{L}_Y=Y^N_{ij}\bar l_i\tilde H_{1}N_{Rj}+Y^E_{ij}\bar l_iH_2E_{Rj}+h.c. 
\end{equation} 
The left handed doublet $l_i$ is a singlet under $U(1)_H$. For the right handed fields, we select for $N_{Ri}$ to have the charge of $H_1(x)$ and for $E_{Ri}$ to have opposite charge of $H_2(x)$. This assignment guarantees lepton flavor conservation. There may be another choice with the couplings violating lepton flavor but the small masses of neutrinos suppress these effects.

The cancellation of anomalies in the 2HDM is a more complicated problem. However there is still enough freedom to select the new charges and, perhaps, introduce additional particles for the cancellation of triangle anomalies~\cite{Ko:2012hd,Ko:2015fxa}.

The models discussed in this section have small neutrino masses with their couplings to neutral scalar particles suppressed relative to the Fermi interaction. In the model of section~\ref{sub31}, the masses are arbitrary parameters to be determined by experiments.  I used the experimental values to show that the amplitudes mediated by the scalar $\phi$ are also small (equation~\eqref{eq:yui15}). 

For the type-I model of section~\ref{sub32} all standard model fermions receive Dirac masses from the Higgs $H_1$. Introducing an additional scalar $\phi$s, being a singlet under $U(1)_H$, makes possible the generation of Majorana masses. The mass matrix now is a seesaw mass matrix of type-I, yielding

\begin{equation}
m_\nu=-m^T_DM^{-1}_Rm_D.\label{eq:las}
\end{equation}
Masses for active neutrinos are inversely proportional to $M_R$ and I assumed $M_R\gg m_D$ in order to secure small masses. Assuming in addition that charged lepton masses are generated by Dirac couplings and that $M_{H_1}$ is the mass of the standard Higgs, FCNC are suppressed relative to the Fermi interaction. There are many other extensions of the standard model with a seesaw mechanism and lepton-flavor changing couplings, for them I refer to a recent publication~\cite{Abada:2021zcm} and the references given there. It appears that whenever two or more scalar particles contribute to the masses of neutral leptons there are lepton-flavor changing couplings and interactions, provided that at least two of the scalars are not isosinglets of $SU(2)_L$.

\section{Summary}
Lepton flavors are conserved in charged lepton decays and interactions, but broken in neutrino oscillations. This motivated the present article in order to investigate and compare whether the rules developed for quarks apply also to leptons. In the introduction we listed the three requirements developed for quarks~\cite{Paschos:1976ay,Glashow:1976nt,Paige:1977nz}. Requirements~\eqref{eq:yui1} and~\eqref{eq:yui2} are satisfied also for leptons, provided that leptons of the same electric charge and chirality have the same values of weak $T^2_L$ and $T_{L3}$. There remains to examine the third requirement that "neutral Higgs couplings conserve lepton flavors to lowest order". 

In section~2 we reviewed the situation in the standard model where one neutral Higgs particle survives. The mass matrix for neutrinos and their couplings to the Higgs become diagonal at the same time. Lepton flavors are conserved in neutral currents because the physical neutrinos are unitary superpositions of flavor states.

In the third section we discuss properties of leptons beyond the standard model. An interesting model has two scalar fields. A Higgs doublet $H(x)$ and a singlet. Bilinears of the scalar fields can couple to fermion pairs. However, the coupling of neutrinos to each scalar is proportional to the mass matrix, thus lepton flavors are conserved. The last case we considered are extended models with a multiplicative $U(1)_H$ factor and two Higgs doublets where solutions to the flavor conservation problem and the cancellation of anomalies are available.

\begin{acknowledgement}
I wish to thank Dr. I. Alikhanov for help in preparation of the article. This is a report for work in progress and may develop into a larger article.
\end{acknowledgement}


%







\end{document}